\documentstyle[12pt]{article}
\include{matex}
\begin{document}
\begin{titlepage}
\begin{center}
\vskip 0.3cm
{\Large \bf R Parity Can Spontaneously Break}
\vskip .3cm
{\large \bf J. C. Rom\~ao}$^1$ \\
{\it Centro de F\'{\i}sica da Mat\'eria Condensada, INIC\\
Av. Prof. Gama Pinto, 2 - 1699 Lisboa Codex, PORTUGAL}\\
{\large \bf C. A. Santos} \\
{\it Faculdade de Engenharia, Dept. de Engenharia Civil\\
Secc\~{a}o de Matem\'{a}tica e F\'{\i}sica, Rua dos Bragas,
4099 Porto Codex, PORTUGAL}\\
and \\
{\large \bf J. W. F. Valle}$^2$ \\
\vs .2cm
{\it Instituto de F\'{\i}sica Corpuscular - C.S.I.C.\\
Departament de F\'isica Te\`orica, Universitat de Val\`encia\\
46100 Burjassot, Val\`encia, SPAIN}\\
\end{center}
\vfill
$^1$ {\it Bitnet ROMAO@PTIFM\\}
$^2$ {\it Bitnet VALLE@EVALUN11 - Decnet 16444::VALLE\\}
\end{titlepage}
In a recent paper \cite{Chaichian} Chaichian and Smilga
criticized the recently proposed supersymmetric extension
of the Standard Model suggested in ref. \cite{MASI} and
defined by the superpotential
\beq
h_u u^c Q H_u + h_d d^c Q H_d + h_e e^c \ell H_d +
\hat{\mu} H_u H_d +
(h_0 H_u H_d - \epsilon^2 ) \Phi +
h_{\nu} \nu^c \ell H_u + h \Phi \nu^c S
%% + M \nu^c S + M_\Phi \Phi \Phi + \lambda \Phi^3
\label{P}
\eeq
For simplicity we consider an effective one generation
model, where the only nonzero entries of the coupling
matrices ${h_{\nu}}_{ij}$ and $h_{ij}$ are those of the
third generation (we have checked, however, that the
flavour violating terms can cause \nt to decay via
Majoron emission with a cosmologically acceptable
lifetime \cite{fae}).
The bilinear $H_u H_d$ term is kept for more flexibility
in obeying LEP constraints (not taken into account in
\cite{Chaichian}). All terms that do not play a role
for our present considerations are neglected. The
scalar potential along neutral directions is taken as
\bea
\label{V}
V_{total}  =
	\abs {h \Phi \tilde{S} + h_{\nu} \tilde{\nu} H_u }^2 +
	\abs{h_0 \Phi H_u + \hat{\mu} H_u}^2 + \\\nonumber
	\abs{h \Phi \tilde{\nu^c}}^2 +
	\abs{- h_0 \Phi H_d  - \hat{\mu} H_d +
	h_{\nu} \tilde{\nu} \tilde{\nu^c} }^2+
	\abs{- h_0 H_u H_d + h \tilde{\nu^c} \tilde{S} - \epsilon^2}^2 +
	\abs{h_{\nu} \tilde{\nu^c} H_u}^2\\\nonumber
+ \tilde{m}_0 \left[-A ( - h \Phi \tilde{\nu^c} \tilde{S}
+ h_0 \Phi H_u H_d - h_{\nu} \tilde{\nu} H_u \tilde{\nu^c} )
+ (1-A) \hat{\mu} H_u H_d
+ (2-A) \epsilon^2 \Phi + h.c. \right]\\\nonumber
	+ \sum_{i} \tilde{m}_i^2 \abs{z_i}^2
+ \alpha ( \abs{H_u}^2 - \abs{H_d}^2 - \abs{\tilde{\nu}}^2)^2
\eea
where $z_i$ denotes any neutral scalar field in the theory
and $\tilde{m}_i$ are the corresponding soft-breaking masses,
$\alpha=\frac{g^2 + {g'}^2}{8}$, and A is the cubic soft
breaking parameter in units of the gravitino mass $\tilde{m}_0$.
We denote the vacuum expectation values (VEVS) by
$v_u = \VEV {H_u}$, $v_d = \VEV {H_d}$,
$v_R = \VEV {\tilde{\nu^c}_{\tau}}$,
$v_L = \VEV {\tilde{\nu}_{\tau}}$,
$v_S = \VEV {\tilde{S_{\tau}}}$ and
$v_F = \VEV {\Phi}$.

In ref. \cite{Chaichian} it is claimed that {\it the extremum
of the potential found in this model is a saddle point,
while in the true minimum R-parity is not broken}.
Without doing any further calculations they also claim that
{\it the problem to construct a viable model involving
spontaneous R-parity breaking remains open}.

The goal of this comment is to refute these claims
and clarify the situation. The no-go "theorem" of
\cite{Chaichian} relies on two assumptions:
(i) all soft-breaking scalar masses are equal to a common
value at the weak scale, and (ii) $h_\nu$ and $v_L$ can be
neglected in the minimization of the scalar potential.
Although (i) is expected to hold at some ultrahigh
unification scale, it is not so at the electroweak scale,
relevant for us. For small $h_\nu$ (i) would lead to
$\tan \beta = \frac{v_u}{v_d} \simeq 1$, while we find
many solutions with $\tan\beta >1$. However, the most
crucial flaw in \cite{Chaichian} is the uncritical use
of assumption (ii). Indeed, if $h_\nu = 0$ and $v_L=0$ the
theory always conserves R parity {\it irrespective of whether
or not nonzero VEVS $v_R$ and $v_S$ are induced.} It easy to
realize this from \eq{P} because in this case we can assign
a positive $R_p$ for the scalars in the chiral superfields
$\nu^c$ and $S$ so that R parity can $never$ break,
irrespective of the potential! Thus this idealized
limit is unsuited to study the dynamics of R parity
breaking. Failure to realize this point is the basis
of the no-go "theorem" in \cite{Chaichian}.

The minimization of the potential in \eq{V} was given in
ref. \cite{pot3}. There we describe in detail the method
and criteria applied. Here we illustrate the main points.
We assume that colour and electric charge are unbroken,
in analogy with what has been verified to hold, for suitable
parameters, in the corresponding $R_p$ conserving model \cite{BFS}.
Instead of directly solving the extremization equations that follow
from \eq{V} with respect to the VEVS, we evaluate the squared-mass
matrices of the neutral scalars derived from \eq{V} and study
their positivity in parameter space. This also allows us
to systematically discriminate against trivial solutions which
are either unphysical (no electroweak breaking) or uninteresting
(no R parity breaking), as well as against solutions that violate
experimental limits. We find a large space of possible minima where
both \21 and R parity break. For illustration, we select that defined
by the following parameters:
$h_\nu = 8.59 \times 10^{-3}$,
$h=-0.351$,
$h_0=0.140$,
$A=1.196$,
$\epsilon^2 = -3.715 \times 10^{5} \: GeV^2$,
$\tilde{m}_0 =355.6 \: GeV$,
$\hat{\mu}=-23.8 \: GeV$,
$\mu_{eff} = \hat{\mu} + h_0 v_F = 94.1\:  GeV$,
$\tilde{m}_d=426.9 \: GeV$,   	$\tilde{m}_u=205.0 \: GeV$,
$\tilde{m}_L=386.8 \: GeV$,	$\tilde{m}_F=355.7 \: GeV$,
$\tilde{m}_R=409.7 \: GeV$,	$\tilde{m}_S=409.7 \: GeV$,
and the corresponding VEVS
$v_d=81.65 \: GeV$,
$v_u=153.77 \: GeV$,
$v_L=-35.9 \: MeV$,
$v_R= v_S=50.00 \: GeV$ and
$v_F=840.89 \: GeV$.
We checked that this minimum is not in conflict with
present LEP limits \cite{LEP1} and that it lies indeed
lower than the trivial minima discussed above. The figure
shows how the potential behaves as $v_L, v_R$ vary around
this minimum, illustrating the presence of a strong hierarchy
between $v_L$ and $v_R$, as required in order to avoid excessive
stellar cooling rates and in order to explain the solar
\neu deficit by resonant \ne to \nm conversions \cite{RPMSW}.

In conclusion, for a wide range of suitably chosen
parameters it is energetically favourable to spontaneously
break both R parity and electroweak symmetries in the model
of \eq{P} and \eq{V}, thus refuting the claims of ref.
\cite{Chaichian}. The characteristic $R_p$ breaking scale
lies in the range $\sim 10\:GeV-1\:TeV$. Implications of
spontaneous $R_p$ violation were considered in \cite{ROMA}.

This work was supported by CICYT. One of us (C.A.S.)
thanks Univ. do Porto for the use of their computers.
\newpage
\section*{Figure Captions}
{\bf Fig. 1}:
Plot of $[V(v_L,v_R)-V_{min}]/V_{min}$ around the minimum
discussed in the text, as a function of the $R_p$ violating
VEVS $v_L$ and $v_R$.
\newpage
\bibliographystyle{ansrt}
\bibliography{biblio}
\end{document}